\documentclass[aps,prc,preprint,superscriptaddress,showpacs]{revtex4}

\usepackage{graphicx}
\usepackage{textcomp}
\usepackage{amsfonts}
\usepackage{amssymb}
\usepackage{bm}
\usepackage[dvips]{color}
\def\ket#1{\vert \,#1 \,\rangle}
\def\bra#1{\langle \,#1 \,\vert}
\def\Kbar{\overline{\!K}}

\begin{document}

\title{Faddeev-Chiral Unitary Approach to the $K^- d$ scattering length}

\author{T. Mizutani}
\email[]{mizutani@vt.edu}
\affiliation{Department of Physics, Virginia Polytechnic Institute 
and State University,\\
Blacksburg, VA 24061, USA}
\affiliation{Theory  Center,  Thomas Jefferson National Accelerator Facility, 
Newport News, VA 23606,  USA}

\author{C. Fayard}
\email[]{c.fayard@ipnl.in2p3.fr}
\affiliation{Institut de Physique Nucl\'eaire de Lyon, IN2P3-CNRS, \\
Universit\'e Claude Bernard, F-69622 Villeurbanne cedex, France }

\author{B. Saghai}
\email[]{bijan.saghai@cea.fr}
\affiliation{Institut de Recherche sur les
lois Fondamentales de l'Univers, DSM/Irfu, CEA/Saclay, F-91191 Gif-sur-Yvette, 
France}

\author{K. Tsushima}
\email[]{kazuo.tsushima@gmail.com}
\affiliation{CSSM, School of Chemistry and Physics, The University of Adelaide,
SA 5005, Australia} 
\date{\today}

\begin{abstract}
Our earlier Faddeev three-body study in the $K^-$-deuteron scattering length, $A_{K^-d}$, is revisited here 
in the light of the recent developments in two fronts:  
{\it (i)} the improved chiral unitary approach to the theoretical description of the coupled $\Kbar N$ related 
channels at low energies, and 
{\it (ii)} the new and improved measurement from SIDDHARTA Collaboration of the strong interaction energy
shift and width in the lowest $K^-$-hydrogen atomic level. 
Those two, in combination, have allowed us to produced a reliable two-body input to the three-body calculation.
All available low-energy $K^-p$ observables are well reproduced and predictions for the $\Kbar N$ scattering 
lengths and amplitudes, $(\pi \Sigma)^\circ$ invariant-mass spectra, as well as for $A_{K^-d}$ are put forward 
and compared with results from other sources. 
The findings of the present work are expected to be useful in interpreting the forthcoming data from CLAS, HADES, 
LEPS and SIDDHARTA Collaborations.
\end{abstract}

\maketitle

%
\section{Introduction} 
There has been a fair amount of recent interest in the low energy interaction of the $\Kbar$ with the nucleon,  
the few nucleon systems, as well as nuclear matter.   
From the point of view of  possible quasi-bound states, particularly the $\Kbar NN$, quite a 
few model discussions have been made during the last decade.   
Relevant references  may be found in 
Refs.~\cite{Kaiser:1995eg,Oset:1997it,Bahaoui:2002yc,Bahaoui:2003xb,Yamazaki:2002uh,Akaishi:2002bg,Yamazaki:2007cs,
Shevchenko:2006xy,Shevchenko:2007zz,Ikeda:2007nz,Ikeda:2008ub,Dote:2007rk,Dote:2008in,Dote:2008hw,Ikeda:2010tk,
Ikeda:2011pi,Ikeda:2012au,Bayar:2011qj,Oset:2011cf,Bayar:2012hn,Sekihara:2012wj,Gazda:2012ic,Mai:2012dt,Hyodo:2011ur}.

The essential ingredient is, of course, in the two-body $\Kbar N$  system.      
From late 70's, during about two decades, data coming from kaonic hydrogen atom created significant
confusion in this realm, since they were impossible to reconcile with available theoretical approaches 
(see e.g. Ref.~\cite{Martin:1980qe}).
Finally in 1997, new data on the strong interaction level shift ($\Delta E_{1s}$) and width ($\Gamma_{1s}$) 
of the $K^-$-hydrogen atomic level become available from KEK~\cite{Iwasaki:1997wf,Ito:1998yi} and was found, 
for the first time, to be consistent with the low energy $K^-p$ scattering data. 
In 2005, a new measurement of $\Delta E_{1s}$ and $\Gamma_{1s}$ became available from the DEAR 
Collaboration~\cite{Beer:2005qi}, but statistically that appeared to be mutually exclusive with the earlier KEK 
result.
Finally, in 2011 the SIDDHARTA Collaboration released~\cite{Bazzi:2011zj} a more precise measurement of 
$\Delta E_{1s}$ and $\Gamma_{1s}$. 
To shed a light on the situation with respect to the three sets of data, we recall below the central values and 
associated total uncertainties $\delta _{tot}=\sqrt{\delta^2 _{stat}+\delta^2 _{sys}}$ 
\begin{eqnarray}
\Delta E_{1s}^{KEK} &=& -323 \pm 64 ~eV~;~\Gamma_{1s}^{KEK}= 470 \pm 231~eV,\\
\Delta E_{1s}^{DEAR}&=& -193 \pm 37~eV~;~\Gamma_{1s}^{DEAR}= 249 \pm 115~eV,\\
\Delta E_{1s}^{SIDD} &=&  -283 \pm 36 ~eV~;~\Gamma_{1s}^{SIDD}= 541 \pm 92~eV.
\label{eq:KDS}             
\end{eqnarray} 
Then, a few comments are in order: 

  {\it (i)} 
Comparing DEAR and SIDDHARTA data shows almost identical uncertainties on $\Delta E_{1s}$ and an improvement of 
about 20\% on $\Gamma_{1s}$, however the discrepancies between the central values come out to be about 
$2.5\sigma$ for both $\Delta E_{1s}$ and $\Gamma_{1s}$.
 
 {\it (ii)}  
Discrepancies in the central values between KEK and SIDDHARTA data show agreements within $1\sigma$ for both 
$\Delta E_{1s}$ and $\Gamma_{1s}$.

  {\it (iii)} 
SIDDHARTA data improve significantly the precision compared to those reported by KEK, namely, total uncertainties 
go down from 20\% to 13\% for $\Delta E_{1s}$ and from 49\% to 17\% for $\Gamma_{1s}$.

In spite of ups and downs in experimental results, on the theoretical side, inspired by an earlier 
work~\cite{Kaiser:1995eg}, an advanced description of the low energy coupled $\Kbar N$ system became 
available~\cite{Oset:1997it}, which was based on the non-linear chiral Lagrangian for the interaction of the octet 
pseudoscalar mesons and octet baryons; hereafter we refer to this type of approaches~\cite{Oller:2000fj,Oller:2000ma,
Lutz:2001yb,Oset:2001cn,Hyodo:2002pk,Jido:2003cb} as the {\it  Chiral Perturbation Theory} ($\chi PT$).

New and improved $\chi PT$ calculations of the coupled $\Kbar N$ channels observables were then developed~\cite{Borasoy:2004kk,Borasoy:2005ie,Borasoy:2006sr,Oller:2005ig,Oller:2006ss,Oller:2006jw}, 
taking up to the next-to-leading order (NLO) terms in the chiral expansion used in the driving terms of the 
scattering equation.  
Borasoy {\it et al.}~\cite{Borasoy:2004kk,Borasoy:2005ie,Borasoy:2006sr} showed that the inclusion of the 
DEAR data made some postdicted cross sections deviate significantly from the data. 
Hereafter the paper by Borasoy, Nissler and Weise~\cite{Borasoy:2005ie} is referred to as {\bf BNW}.
Still within another $\chi PT$ approach~\cite{Oller:2005ig,Oller:2006jw}, two types of solution amplitudes were 
found: one ($A^+_4$) consistent with the DEAR data, but the other ($B^+_4$) not. 
A more recent work~\cite{Cieply:2007nv,Cieply:2009ea} used a chirally motivated separable model to study the 
coupled $\Kbar N$ channels, producing somewhat lower $K^-p$ elastic cross sections compared to the data, and the 
atomic level width came out to be quite larger in magnitude than the DEAR data, although the latter 
was used to constrain the fit. 

Finally, with regard to works incorporating the SIDDHARTA data, there are two new and improved $\chi PT$ 
calculation of the $\Kbar N$ amplitudes~\cite{Ikeda:2011pi,Mai:2012dt}; to be discussed later.  

During that period, $K^-$d scattering length was investigated by several authors, though no data are still available.  
The $\chi PT$ was exploited~\cite{Kamalov:2000iy} to perform a three-body calculation of  $A_{K^- d}$, within 
the fixed center approximation (FCA) in the input $\Kbar N$ amplitudes.  
In the presence of a two-body resonance near threshold (i.e. $\Lambda (1405)$) this approximation had been known 
to lack accuracy.  
We  performed a relativistic three-body calculation of this quantity~\cite{Bahaoui:2003xb} that automatically took 
into account such effects as nucleon-binding, target recoil, intermediate nucleon-hyperon interactions, etc.   
Also we demonstrated~\cite{Bahaoui:2002yc,Bahaoui:2003xb} the importance of retaining the deuteron $D$-state, estimated 
other possible effects not included in the calculations such as, three-particle forces, Coulomb interaction, etc. 
Our model prediction was $A_{K^-d}=(-1.80 + i 1.55)$ {\it fm} with an uncertainty of about $\pm 10\%$.   
Later, a work~\cite{Sibirtsev:2004kk} came out in which an ingenious method was devised in extracting the 
$\Kbar^\circ d$ scattering length from the  reaction $pp \to \Kbar^\circ K^+d$~\cite{Kleber:2003kx}.  
The result appeared to prefer a smaller value in magnitude both for the real and imaginary parts: it looked as if our model 
result were off by two to three standard deviations.  
However, in view of some assumptions and approximations adopted in Ref.~\cite{Sibirtsev:2004kk} (such as the scattering length 
approximation to the final state enhancement factor which might well vary rapidly due to the $\Lambda (1405)$),  
their prediction might be of semi-quantitative nature (we note that the quoted values from our model prediction in that 
work were incorrect).     

During the DEAR era, Meissner {\it et al.}~\cite{Meissner:2006gx} put forward the ranges of allowed values for scattering 
lengths $a_{\Kbar N}$  and $A_{K^-d}$ by making use of the FCA. 
With the value of $a_{K^-p}$ extracted from the DEAR experiment, almost all the existing model results  for $A_{K^-d}$ 
were shown to stay outside the determined limits. 
Nevertheless, the corresponding limits on the range extracted from the KEK data happened to be far more accommodating.   

At the three-body level, an early work~\cite{Meissner:2006gx} was improved and came up with results~\cite{Doring:2011xc} 
less restrictive than with the input from the DEAR data for $A_{K^-d}$.  
The inclusion of the recoil effect to FCA~\cite{Baru:2009tx} is expected to modify the prediction 
somewhat for the better.   
In a recent work~\cite{Shevchenko:2012wm} on $A_{K^-d}$ a standard separable interaction~\cite{Bahaoui:1990da} was adopted, 
but with two-body two-channel ($\Kbar N -\pi \Sigma$) potentials with a pole (to mimic the $\Lambda (1405)$) fitted to 
the principal $K^-p$ initiated channels, and the SIDDHARTA data.  
This is a refinement to a somewhat earlier work by the same author~\cite{Shevchenko:2011ce}.  
We note yet another piece of recent work~\cite{Faber:2010iw} in which the SIDDHARTA result was incorporated in a FCA 
calculation of $A_{K^-d}$ as well as the corresponding $K^-$-deuteron $p$-wave scattering volume to predict the 
$p$-wave atomic energy shift and width in the $K^-$-deuterium, where two sets of very distinct solutions were reported.

Following recent findings, in the present paper we revisit our previous work~\cite{Bahaoui:2002yc,Bahaoui:2003xb} 
within an improved $\chi PT$ for the coupled channel $\Kbar N$ amplitudes, using them to obtain $A_{K^-d}$.  
 
The organization of the present paper is as follows: Section~\ref{sec:two-body} describes the $\chi PT$ 
approach to the coupled $\Kbar N$ channels, which is the primary input to our calculation of $A_{K^-d}$, 
Section ~\ref{sec:theory} embodies a brief description of the three-body calculation we have 
adopted~\cite{Bahaoui:2003xb}.
Numerical method is outlined in Section~\ref{sec:practical-cal} and our results reported and discussed in 
Section~\ref{sec:results}.   
Finally, Section~\ref{sec:conclu} serves as drawing our conclusions.
We relegate to Appendix~\ref{apdx:A} some of the useful formulae for Section~\ref{sec:two-body}.    
%
%
\section{Two-body channels}
\label{sec:two-body}
Since the nucleon-nucleon ($N N$) two-body interaction that we have adopted in the present work is the same as that 
used in Refs.~\cite{Bahaoui:2002yc,Bahaoui:2003xb}, this section is devoted to the coupled $\Kbar N$ channel interactions 
that are the primary two-body ingredient to the three-body approach. 

For the two-body $\Kbar N$ channels, we adopted a chiral interaction Lagrangian which contains the meson 
$\phi \equiv(\pi, K, \eta)$ and baryon $B\equiv N,~\Lambda,~\Sigma,~\Xi$ fields, as reported below.   
\subsection{$\Kbar N$ Interactions}
\label{sec:kbarn_inter}
%
Coupled channel equations determine the reaction among various two-body meson-baryon channels.  For our 
present study these are $\Kbar N$, $\pi Y$, $\eta Y$, and $K \Xi$ with different charge states.   
By using $i, \  j, \ k, ...$, as channel indices, the coupled Bethe-Salpeter equations for the $t$-matrices  
for the scattering process $i \to j$ reads  
\begin{equation}
T_{ij} =  V_{ij} + \sum_{k} V_{ik} G^k_0 T_{kj},
\label{eq:tijls}             
\end{equation}  
where $V_{ij}$ is the transition potential (or the driving term) taken from an effective chiral Lagrangian to be 
discussed below, and $G^k_0$ is the free meson-baryon propagator for the intermediate channel $k$. 
We note here that implicit in the above expression are that {\it (i)} although not essential, the meson-baryon 
systems are in the two-body center of mass (c.m.) frame, and {\it (ii)} the integration is performed over the off-shell 
four momentum  associated with channel $k$. 
It is also possible to write the above set of equations collectively in a simple matrix form
\begin{equation}
\widetilde T= \widetilde V  + \widetilde V \widetilde G \widetilde T ,
\label{eq:tls}
\end{equation}
where $\widetilde V =\{V_{ij}\}$  is the matrix of driving terms, and similarly for $\widetilde T$, while
$\widetilde G$ is a diagonal matrix with elements $G^k_0$ in the diagonal. 

Next, in a standard $\chi PT$ approach one makes an {\it on-shell} ansatz ~\cite{Oset:1997it} in which
both $\widetilde V$ and $\widetilde T$ are put fully {\it on-shell} with respect to the initial c.m. channel energy, 
$W \equiv \sqrt{s}$, with $s$ being the Mandelstam $s$-variable for all the channels involved. 
Then this ansatz renders the above equation into an algebraic one for which the solution becomes
 \begin{equation}
 \widetilde T = 
 \left [ 1 - \widetilde V  \widetilde G(s) \right ]^{-1} \widetilde V , 
 \label{eq: solT}
 \end{equation}
where the diagonal elements of $\widetilde G(s)$ are now the {\it off-shell momentum} integrated $G^k_0$(s), often
called {\it scalar loops} and are functions of only $s$. 
Note that our sign convention for $\widetilde V$ and $\widetilde T$ is the opposite to the one adopted in 
Refs.~\cite{Borasoy:2004kk,Borasoy:2005ie}.   
A more detailed account is given in Appendix~\ref{apdx:A}.   
  
Now we need to specify the fully on-shell $\widetilde V$ (or $V_{ij}$). 
As has become standard by now, one obtains these driving terms from a Lagrangian with non-linear realization of chiral 
symmetry. 
We refer the reader to Refs.~\cite{Borasoy:2004kk,Borasoy:2005ie} whose notation and description we will closely follow 
in our subsequent description.
  
\begin{figure}[ht]
\includegraphics[width=0.78\linewidth]{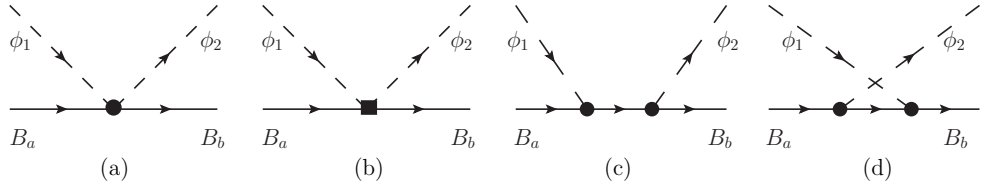}
\vspace{-12.0cm} 
\caption{\footnotesize Diagrams for meson-baryon scattering: seagull or the Weinberg-Tomozawa interaction (a), 
contact interactions or the Next-to-Leading Order (b), direct $s$-channel (c), and crossed $u$-channel Born terms.
Dashed and solid lines are for mesons and baryons, respectively.  
}
\label{fig:Vterms}
\end{figure}

Four terms from two primary sources are included for $\widetilde V$, Fig.~\ref{fig:Vterms}, as follows:
 
{\it (i)} The leading order contact term in the Lagrangian providing the commonly called Weinberg-Tomozawa (WT) 
leading order contact term $\widetilde V^{(a)}$, two Born terms: 
direct $s$-channel $\widetilde V^{(c)}$ and crossed $u$-channel $\widetilde V^{(d)}$, which contain two 
axial vector coupling constants $D$ and $F$. 
As in BNW, we have adopted the frequently used central values~\cite{Close:1993mv}, namely, $D=0.80$ and $F=0.46$, 
subject to constraint by the nucleon axial-vector coupling constant in the chiral limit, i.e., $D+F=g_A =1.26$.
   
{\it (ii)} The NLO term $\widetilde V^{(b)}$ is of the second chiral order $\mathcal{O}(p^2)$, and contains a chiral symmetry 
breaking measure ($B_0$) arising from the chiral condensate, as well as $u$, $d$ and $s$ current quark masses.  
The dependence on those quantities is eventually converted to that on the physical meson masses,
using the Gell-Mann-Oakes-Renner relation for the Goldstone boson masses.  
In addition, for our current objective, this part of the interaction depends on seven 
{\it low energy constants}~\cite{Kaiser:1995eg}:  
$b_0$, $b_D$, $b_F$, $d_1$, $d_2$, $d_3$ and $d_4$, which should be determined from fits to relevant physical observables.  
As for the first three parameters which are determined from our present fit to available low energy cross sections,  
we will call them the {\it renormalized} $b$ parameters: $\bar b_0$, $\bar b_D$ and $\bar b_F$.  
The reason for this may be found in Refs.~\cite{Ikeda:2011pi,Ikeda:2012au}, and in our later discussion.  
For convenience, we also introduce different combinations of driving terms, called $WT$-, $c$-, $s$-, and $u$-model as follows:
%
%
\begin{equation}
\label{eq:Vs}
\widetilde V_{WT} \equiv \widetilde V^{(a)}, 
\ \widetilde V_c \equiv \widetilde V^{(a)} + \widetilde V^{(b)}, 
\  \widetilde  V_s   \equiv  \widetilde V^{(a)} 
+ \widetilde V^{(b)} +  \widetilde V^{(c)}, 
\ \widetilde V_u \equiv \widetilde V^{(a)} + \widetilde V^{(b)} +  
\widetilde V^{(c)} + \widetilde V^{(d)},
\end{equation}
so, for example, the $s$-model contains WT, NLO and $s$-pole terms.  

%

For the present three-body study with total strangeness of the input meson-baryon system $S=-1$, we need two separate 
sets of  coupled channels among various two-body physical particle states:  
%
%
\begin{equation}
\label{eq:ch1-8}
K^-~p~\to~
K^-p~,~ \Kbar^\circ n~, ~\Lambda \pi^\circ~,~ \Sigma^+ 
\pi^-~,~\Sigma^\circ \pi^\circ~,
~\Sigma^-\pi^+~, ~ \Lambda\eta~, ~ \Sigma^\circ\eta~,
~\left [K^+\Xi^-\right ],~\left [K^\circ\Xi^\circ\right ],
\end{equation}
\begin{equation}
\label{eq:ch1-5}
K^- n \to K^- 
n~,~\Lambda\pi^-~,~\Sigma^\circ\pi^-~,~\Sigma^-\pi^\circ~,~\Sigma^-\eta~,
~\left [K^\circ \Xi^-\right ].
\end{equation}
The first set contains ten channels ($N_{K^-p}=10$) and the second one six ($N_{K^-n}=6$).   
Thus  for the first set, Eq.~(\ref{eq:ch1-8}), channels indices $i,~j$~for $V_{ij}$ run from $1$ to $10$.   
We note that  the square bracketed channels in the two equations above were {\it not} included in our 2002/2003 
works~\cite{Bahaoui:2002yc,Bahaoui:2003xb} although most of the later works have retained them conventionally.   
As explained later, in the present study we began with the [$N_{K^-p}=10, N_{K^-n}=6$] scheme, then based on the 
statistical fit to data, eventually went back to the earlier choice of [$N_{K^-p}=8, N_{K^-n}=5$].    

Having specified the channels, the concrete form of the driving term may now be identified; 
Refs.~\cite{Borasoy:2005ie,Ikeda:2012au} provide all the necessary information; see also Sec. III 3 of 
Ref.~\cite{Hyodo:2011ur} (note the difference in the overall sign and in spinor normalization in these references). 
These driving terms may need to be spin averaged and projected onto the $s$-wave orbital state before fed into the 
scattering equation.    
In these expressions of the driving terms, the weak decay constant $f$ enters everywhere.   
Recall that due to the on-shell ansatz~\cite{Oset:1997it}, its value is expected to be different either from the 
one in chiral limit, or the physical ones for pion, kaon, or  $\eta$ decays.   
For our objective its value is determined by fits to the scattering data whereas in  recent works~\cite{Ikeda:2011pi,Ikeda:2012au} 
$f_{\pi}$ was  fixed at the physical value and $f_K$, $f_\eta$ were floated around their physical values and determined by fits. 
 
Once the driving terms are specified, the last quantity we need to take care of,  before performing statistical fits to 
determine the scattering $t$-matrix $\widetilde T$, is $\widetilde G(s)$ whose $j^{th}$ diagonal matrix element is a scalar loop,
\begin{equation}
\label{eq: loop}
G^{j}(s) = \int\frac{d^4q}{(2\pi)^4} \frac{i}
{\Big[ (p-q)^2 -M^2_j + i\epsilon \Big] \Big[q^2 - m^2_j + i\epsilon \Big]}, 
\end{equation}
where $p^2=s$, and $M_j$ ($m_j$) are the baryon (meson) masses in the $j^{th}$ channel.  
This is a divergent integral, made finite by dimensional regularization which introduces subtraction constants 
$a_j$, $j=1,~2,~...$.  
These constants are also to be fixed by fits to the scattering data.  
Since a scalar loop is characterized only by masses of the particles involved, assuming isospin symmetry to reduce the 
number of these subtraction constants may be quite relevant, so for the number of channels $N_{K^-p}=10$ (8) the 
number of $a_j$s is 6 (5).  
For more on the scalar loops, see Appendix~\ref{apdx:A}.     
Here it should be useful to mention a very recent work~\cite{Mai:2012dt}, which is within the $\chi PT$, but without 
the on-shell ansatz.   
Furthermore, {\it (i)} it corresponds to the $c$-model mentioned above, but included more NLO terms (see the reason 
behind this choice in Ref.~\cite{Bruns:2010sv}), 
{\it (ii)} only the first six channels ($N_{K^- p}=6$) are included in the fitting procedure that are open in the 
energy range for the adopted data. 

Before proceeding  to the determination of  $\widetilde T$ using statistical fits to data,  it should be useful to
discuss a couple of items.  

First,  we need to state where the present work is different from our 2002/2003 study~\cite{Bahaoui:2002yc,Bahaoui:2003xb}.  

{\it (i)} Our earlier work~\cite{Bahaoui:2003xb} included only the WT interaction $\widetilde V_{WT}$ in the driving term 
whereas in the present work we go further to include the NLO contribution, so the $c$-model $\widetilde V_{c}$, and also 
the $s$-model $\widetilde V_{s}$.  
In practice we have excluded the $u$-model $\widetilde V_{u}$ for reasons to be stated soon below.   
Note that the form of the WT interaction used then and in the present work are equivalent except for a small contribution 
disregarded in the former.  

{\it (ii)} Our earlier work introduced a global form factor in the driving terms to tame the {\it otherwise divergent} 
integral for loop functions.  
We feel that the dimensional regularization in the present work is more consistent in spirit with respecting chiral symmetry. 

{\it (iii)} Apart from physical masses, the implementation of the $SU(3)$ symmetry breaking was in the different values of 
the meson decay constant in our previous work, but in the values of the subtraction constants in the present study.     

The second item is on  the interaction models we have chosen in the present study as  just stated in {\it (i)} above.  
As mentioned in BNW, upon adopting the on-shell ansatz, and upon projecting onto the $s \ (l=0)$ orbital angular momentum 
state, the $u$-pole contribution $\widetilde V^{(d)}$ develops logarithmic dependence on the Mandelstam variable $s$.  
This type of logarithmic dependence is generally mild and the resultant interaction of lesser importance relative 
to other contributions, except that the  branch cuts from the higher threshold channels extend up to the threshold of some
light meson-baryon channels.  
We should stress that those branch cuts are {\it unphysical} originating from the on-shell ansatz.  
Possible cures may be, 
{\it (a)} as in BNW, to eliminate the singularities by matching 
the amplitude appropriately to some constant 
where relevant, 
{\it (b)} no particular modification to eliminate the branch cuts 
by arguing that they might only affect a less 
dominant amplitude for elastic $\pi^{\circ} \Lambda$ below the 
$\pi \Sigma$ threshold~\cite{Oller:2006ss}, or 
{\it (c)} to adopt a static approximation to baryons in the $u$-pole terms 
as done, for example, in~\cite{Cieply:2009ea}.
  
Here in our work, we simply do not include the $u$-pole contributions in the driving term by acknowledging their
relatively weak contribution as already stated above.   
Thus we have studied just the $c$- and $s$- models.  
Underlining this is the  observation that, as may be easy to understand from the static limit, the $s-$ and $u-$pole 
terms tend to compensate each other.  
Consequently, a {\it properly regulated} $u$-model is expected to provide resultant amplitudes and observables which may well 
come out somewhere  between the corresponding quantities from the $c-$ and $s-$models.  
We should emphasize here that once the coupled two-body $\Kbar N$ amplitudes are fed into the coupled three-body equations,  
the loop momentum integrations go down below the threshold of those two-body amplitudes, so it is essential that the result 
of the three-body calculation not be distorted by any unwanted singularities, hence our choice above!   
Since only BNW have presented separately the cases with $c-$ and $s-$models' results, we were compelled to construct our own
$\Kbar N$ channel amplitudes.

To be complete, in Table  \ref{tab:masses} we list the physical particle masses~\cite{Nakamura:2010zzi} 
used in the present work.  
%
{\squeezetable
\begin{table}[htb]
\squeezetable
\caption{\footnotesize Particle masses (in MeV).}
\label{tab:masses}
\begin{center}
\begin{ruledtabular}
\begin{tabular}{ccccccccccccccc}
  $K^-$       & $K^+$      & $\Kbar^\circ$  & $p$        & $n$ &
  $\pi^-$     & $\pi^+$    & $\pi^\circ$ &
  $\Sigma^-$  & $\Sigma^+$ & $\Sigma^\circ$ & $\Lambda$  & $\eta$ &
  $\Xi^-$    & $\Xi^\circ$ \\  
\hline 
%
  493.68   & 493.68  & 497.65  & 938.27  & 939.57 & 
  139.57   & 139.57  & 134.98  &
  1197.45  & 1189.37 & 1192.64 & 1115.68 & 547.51 &
  1321.31  & 1314.83 \\
\end{tabular}
\end{ruledtabular}
\end{center}
\end{table}
}
\subsection{Data base and extraction of adjustable parameters}   
\label{sec:Data}
Here, we report on the data used to extract the adjustable parameters for our two-body amplitudes.

In the energy range of interest in this work ($P_{K^-} \lesssim 250$ MeV/c), total cross section measurements 
have been performed between 60's and 80's, producing
~\cite{Abr65,Cse65,Sakitt:1965kh,Kittel:1966zz,Kim:1967zz,Mast:1974sx,Mast:1975pv,Bangerter:1980px,Ciborowski:1982et,Evans:1983hz} 
some 90 data points for the following channels:
$K^-p~\to~K^-p,~\Kbar^\circ n,~\Lambda \pi^\circ,~\Sigma^+ \pi^-,~\Sigma^\circ \pi^\circ,~\Sigma^-\pi^+$.
Given that the complete data set shows internal inconsistencies, reduced sets are used in various fitting procedures and the number 
of retained data points is not identical in reported phenomenological investigations.
Here, we have adopted the same approach as in Refs.~\cite{Bahaoui:2002yc,Bahaoui:2003xb,Siegel:1994mb}, 
where 52 cross section data~\cite{Cse65,Sakitt:1965kh,Mast:1974sx,Mast:1975pv,Ciborowski:1982et} were selected.
The method used there was to fit all 90 cross section data, as well as the accurate 
data~\cite{Sakitt:1965kh,Kittel:1966zz,Humphrey:1962zz,Tovee:1971ga,Nowak:1978au} for $K^-p$ reaction
rates at threshold, {\it i.e.},
\begin{eqnarray}
\label{eq:Br}
\gamma & = & 
      {{K^-p \rightarrow \pi^+ \Sigma^-} \over
       {K^-p \rightarrow \pi^- \Sigma^+}} = 2.36 \pm 0.04 ,\\    
R_c & = &  
    {{K^-p \rightarrow \hbox{charged particles}} \over
     {K^-p \rightarrow \hbox{all final states}}} = 0.664 \pm 0.011 
                                                                   ,\\   
R_n & = & 
   {{K^-p \rightarrow \pi^\circ \Lambda} \over
    {K^-p \rightarrow \hbox{all neutral states}}} = 0.189 \pm 0.015 .
\end{eqnarray}

Then, cross section data giving the highest $\chi^2$ were removed and the reduced data base was refitted.
Checking the outcome of several combinations of data sets allowed establishing a consistent enough data base.
In the present work, in addition, the strong interaction level shift ($\Delta E_{1s}$) and width ($\Gamma_{1s}$) 
of the $K^-$-hydrogen atomic level data from the SIDDHARTA Collaboration~\cite{Bazzi:2011zj} were also included 
in the data base, without affecting the acceptable consistency of the kept total cross section data.
Finally, we added to the data base the pion-nucleon sigma term, $\sigma_{\pi N}=35\pm10$ MeV, which brings in a 
loose constraint (we will come back to this issue at the end of this section).
In summary, our data base embodies a total of 58 data points.

Our approach for $N_{K^-p}=8$ contains 13 adjustable parameters, as reported in Table~\ref{tab:param}. 

%
\begin{table}[ht]
\caption{\footnotesize Adjustable parameters in the present work. 
The weak decay constant $f$ is in MeV.
The renormalized NLO parameters ($\bar b_{0}$, $\bar b_{D}$, $\bar b_{F}$)
and the low energy constants 
in the NLO Lagrangian ($d_{1}$, $d_{2}$, $d_{3}$,  $d_{4}$) are in GeV$^{-1}$. 
The subtraction constants $a(\mu)$ are given at $\mu$=1 GeV.}
\label{tab:param}
\begin{center}
\begin{tabular}{lcccc}
  \hline
  \hline
  Parameter && Model $c$ &&  Model $s$  \\  
\hline 
  $f$            && 116.2$\pm$0.7      &&  122.4$\pm$0.9   \\
  $\bar b_0$     && -0.35$\pm$0.01     &&  -0.40$\pm$0.02  \\
  $\bar b_{D}$   && ~0.01$\pm$0.02     &&  -0.08$\pm$0.01  \\
  $\bar b_{F}$   && -0.02$\pm$0.02     &&  -0.04$\pm$0.03  \\
  $d_1$          && -0.17$\pm$0.02     &&  -0.11$\pm$0.03  \\
  $d_2$          && ~0.07$\pm$0.01     &&  ~0.06$\pm$0.01   \\
  $d_3$          && ~0.29$\pm$0.01     &&  ~0.26$\pm$0.02   \\
  $d_4$          && -0.34$\pm$0.01     &&  -0.28$\pm$0.02  \\
  $a _{\Kbar N}$~$(\times 10^{-3})$       && ~1.60$\pm$0.08    
                                          &&  ~1.89$\pm$0.10  \\
  $a _{\pi \Lambda}$~~$(\times 10^{-3})$  && -8.59$\pm$2.33    
                                          &&  ~0.96$\pm$3.29  \\
  $a _{\pi \Sigma}$~~$(\times 10^{-3})$   && ~1.15$\pm$0.33    
                                          &&  -0.05$\pm$0.34  \\
  $a _{\eta \Lambda}$~~$(\times 10^{-3})$ && -2.30$\pm$0.26    
                                          &&  -1.86$\pm$0.34  \\
  $a _{\eta \Sigma}$~~$(\times 10^{-3})$  && ~2.99$\pm$3.91    
                                          &&  -6.35$\pm$4.90  \\
  \hline
  $\chi^2 _{d.o.f.}$  && ~1.22 &&  1.21 \\
  \hline
  \hline
\end{tabular}
\end{center}
\end{table}
%

The uncertainties attributed to our results are those generated by MINUIT, used for minimizations.
Our $c-$ and $s-$models lead both to $\chi^2 _{d.o.f.} \approx$1.2.
As much as the subtraction constants are concerned, taking into account the uncertainties, we observe that 
$a _{\pi \Lambda}$, $a _{\pi \Sigma}$ and $a _{\eta \Sigma}$ are rather poorly determined.

Here, we discuss some issues related to the pion-nucleon sigma term $\sigma_{\pi N}$.
While we want to perform our fit by using direct experimental data only, we are interested in the extent to which 
the constraint from the pion-nucleon sigma term may affect the resulting data fit. 
At leading order the sigma term reads,
\begin{eqnarray}
\label{eq:sigter}
\sigma^0_{\pi N} & = & -2m^2_{\pi} (2b_0 + b_D +b_F).
\end{eqnarray}

It is well known that the pion-nucleon $\sigma$-term, obtained by Gasser {\it et al.}~\cite{Gasser:1990ce},
based on $\pi N$ data analysis and taking into account the current algebra result generated by the quark masses, 
gave $\sigma_{\pi N}$=45$\pm$8 MeV. 
More recent investigations lead either to smaller or larger values with respect to the central one, 45 MeV.
For instance, {\it (i)} a pertubative chiral constituent quark model~\cite{Inoue:2003bk} finds 55 MeV, 
an analysis of $\pi N$ scattering amplitude via chiral perturbation theory~\cite{MartinCamalich:2010fp}
reaches $59 \pm 7$ MeV,
or still a dispersion relations approach~\cite{Hite:2005tg}, using the SAID $\pi N$ phase-shift 
analysis~\cite{Arndt:2003if} gives $\sigma_{\pi N}$=81$\pm$6 MeV, {\it (ii)} on the other hand, lattice QCD 
calculations lead to $39 \pm 4$ MeV~\cite{Durr:2011mp}, 
$38 \pm 12$ MeV~\cite{Bali:2011ks}, $45 \pm 6$ MeV~\cite{Shanahan:2012wh}, while chiral constituent quark 
models~\cite{An:2010wb} predict 31 MeV~\cite{Dahiya:2011uh} or 37 MeV~\cite{An:2012ii}.

In recent works~\cite{Oller:2005ig,Oller:2006jw,Cieply:2007nv,Cieply:2009ea}, including the present one, 
some of the NLO low energy constants (notably $\bar b_0, \bar b_D$ and $\bar b_F$) are constrained by using 
the lowest order terms for meson and/or baryon mass formula (Gell-Mann-Okubo mass formula), as well as the 
pion-nucleon sigma term $\sigma_{\pi N}$ within the context of the chiral perturbation theory ($\chi PT$).

On the other hand, by calculating the corresponding value from several $\chi PT$s for the coupled $\Kbar N$ 
system ~\cite{Borasoy:2005ie,Borasoy:2006sr,Ikeda:2012au}, one finds lower values of 
$\sigma_{\pi N} \approx 15 - 30$ MeV.   
However, as discussed briefly in Ref.~\cite{Ikeda:2012au}, the latter is the result of the fit to scattering 
data by iterating the driving term (with physical hadron masses adopted) which contains those NLO constants 
to infinite orders. 
The extracted values of these constants are then expected to be different from those obtained by fitting baryon
masses perturbatively, see e.g. Ref.~\cite{Ikeda:2012au}.  
So as stated earlier, we denote the former set of the corresponding values as $\bar b_0,  \  \bar b_D, \  \bar b_F$, 
and the corresponding sigma term as $\bar \sigma^{(0)}_{\pi N}$, as in Ref.~\cite{Ikeda:2012au}.
       
In the present work, performing minimizations with no constraint on the sigma term, $\bar \sigma^{(0)}_{\pi N}$ 
was found to fluctuate around $15 \sim 20$ MeV. 
Trying to enforce the sigma term in the region 50 to 85 MeV resulted in clustering near the lowest limit. 
While letting that term vary in the range 25 to 45 MeV, the minimization went smoother, with little effect 
on the $\chi^2$.
So with the loose constraint $25 \leq \bar \sigma^{(0)}_{\pi N} \leq 45$ MeV, we performed the final minimizations. 
The parameters reported in Table~\ref{tab:param} lead to $\bar \sigma^0_{\pi N} =$ 28 MeV and 36 MeV for models 
$c$ and $s$, respectively.

To end this section, we would like to comment about other channels, $K^- p \to \eta \Lambda,~K \Xi$, 
with thresholds beyond the energy range investigated in this work.

With $N_{K^- p}=10$ fit, the last subtraction constant $a_{K\Xi}$ tends to get quite large relative to others,
with sizeable uncertainty.   
This indicates that the two $K\Xi$ channels are very likely not relevant to the low-energy fit, and  the extracted
$a_{K\Xi}$s have no actual substance.  
To test this observation from several $N_{K^- p}=10$ fits, we dropped the last two $K\Xi$ channels in 
Eq.~(\ref{eq:ch1-8}), and ran the effective $N_{K^- p}=10-2=8$ channel models to calculate the corresponding observables. 
The result was found just about $\sim 5 \%$ different from the original $N_{K^- p}=10$ cases.   
To confirm this, we also used $N_{K^- p}=8$ fit results, then  added two extra $K\Xi$ channels from a few $N_{K^- p}=10$ 
fit results, to make an effective $N=8+2=10$ model.  
Again, the latter was found to reproduce the observables calculated in $N_{K^- p}=8$ configuration within $\sim 5 \%$. 
Also with respect to the data set used in the present work, the $N_{K^- p}=8$ minimizations are slightly better than 
the ones with $N_{K^- p}=10$ in $\chi^2_{d.o.f.}$ by about 0.1.  
So we felt justified to adhere to the $N_{K^- p}=8$ model. Therein, we kept $K^- p \to \eta Y$ channels which have lower 
thresholds compared to $K\Xi$ final states, but because of lack of low-energy data, we could only check the smooth rising 
of the cross section close to threshold, matching the lowest energy $K^- p \to \eta \Lambda$ data points reported by 
the Crystal Ball Collaboration~\cite{Starostin:2001zz}.
Finally, recent studies at higher energies can be found e.g. in~\cite{Starostin:2001zz,Liu:2011sw} for $\eta \Lambda$ 
final state and in Refs.~\cite{Oller:2006jw,Sharov:2011xq,Shyam:2011ys} for $K \Xi$ final states. 
\subsection{Results and Discussion}
\label{sec:RD}
In this section we present our results and compare them with the data, namely, cross sections for $K^-p$
initiated reactions, threshold strong channels branching ratios and kaonic hydrogen atom $1s$ level shift.
Then we will report on our predictions for the scattering lengths and amplitudes, as well as for the 
$(\pi \Sigma)^\circ$ invariant-mass spectra. 
%
\subsubsection{Total cross sections}
\label{sec:tcs}
In the low energy range of our current interest, {\it i.e.} $P_K^{lab} \le 250$ MeV/c, the following strong coupled 
channels are open:
\begin{equation}
\label{eq:open}
K^-~p~\to~
K^-p~,~ \Kbar^\circ n~, ~\Lambda \pi^\circ~,~ \Sigma^+ 
\pi^-~,~\Sigma^\circ \pi^\circ~,
~\Sigma^-\pi^+,
\end{equation}
and strongly influenced by the $I=0$ $\Lambda (1405)$ resonance below the $K^-p$ threshold which decays almost exclusively 
to $\pi \Sigma$.
Moreover, while the $\Kbar^\circ n$ has a slightly higher threshold, all the $\pi Y$ channels have lower threshold than that 
for $K^-p$.
%
%
\begin{figure}[ht]
\includegraphics[width=0.9\linewidth]{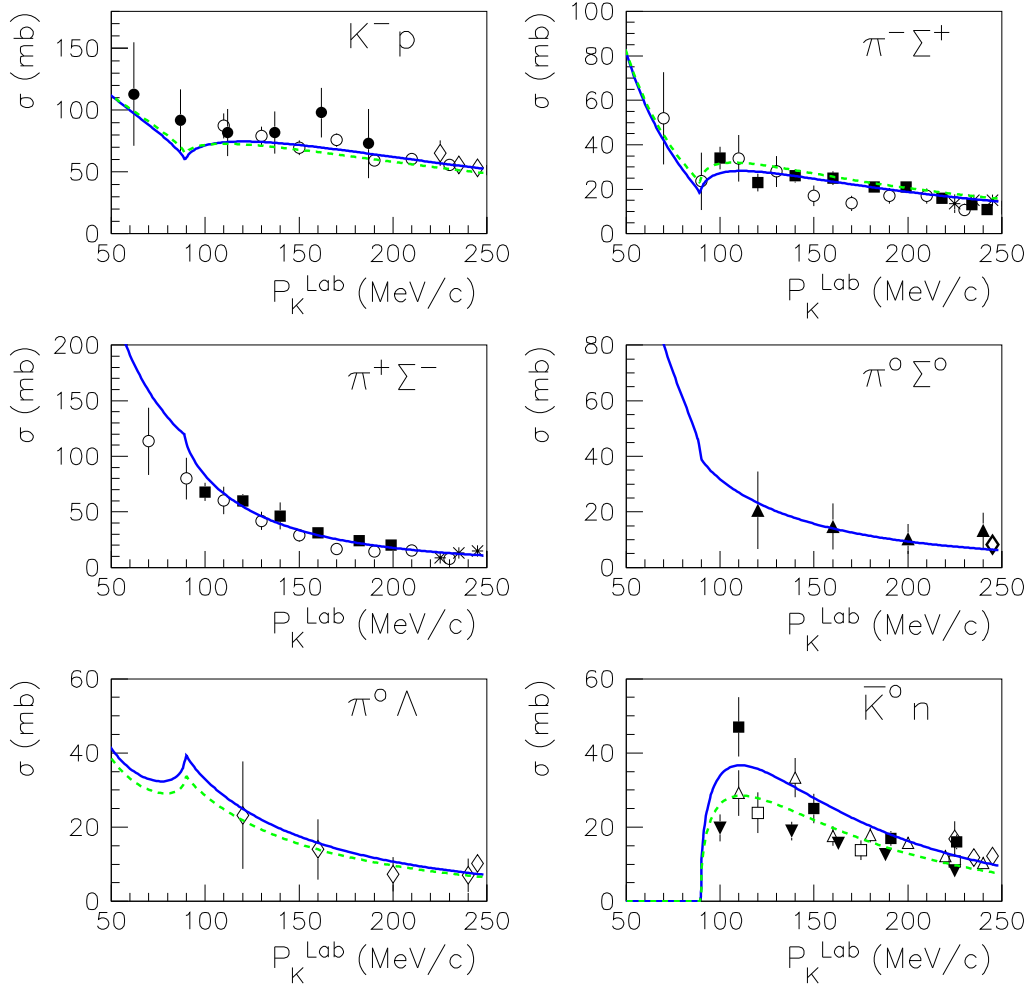}
\caption{(Color online) \footnotesize 
Total cross sections initiated by $K^-p$. 
The solid (blue) curves are our fits and the dashed (green) lines our results for the BNW model~\cite{Borasoy:2005ie}; 
both sets are given for the $c$-model.   
The solid and dashed curves are almost identical for $\pi^+ \Sigma^-$ and $\pi^\circ \Sigma^\circ$.
Experimental data are from Refs.~\cite{Cse65,Sakitt:1965kh,Mast:1974sx,Mast:1975pv,Ciborowski:1982et}.}
\label{fig:xs}
\end{figure}
%

Our $c$-model results for all open strong channels are reported in Fig.~\ref{fig:xs} and 
compared to the data. The $s$-model gives very close values to those of $c$-model, so they are not depicted.

As mentioned above, we have retained 52 data points for those channels. 
The obtained $\chi^2$ per data point ($\chi^2_{d.p.}$) turns out to be 0.98. 
Our model allows reproducing the total cross section data satisfactorily. 
This is also the case for the BNW model, also for the $c$-model (close to the $s$-one).
The only discrepancy between the two models concerns the $\Kbar^\circ n$ final state, due to the respective 
fitted data bases.
\subsubsection{Threshold strong branching ratios and kaonic atom}
\label{sec:tbr}
Table~\ref{tab:res} summarizes theoretical and experimental values for $K^-p$ reaction rates: $\gamma$, $R_c$ 
and $R_n$ (Eqs. (11-13)), as well as $K^-$-hydrogen atomic $1s$ level shift and width.
We get $\chi^2_{d.p.}$=0.99 for the branching ratios and $\chi^2_{d.p.}$=0.21 for the SIDDHARTA kaonic hydrogen 
atom data.

Our $c-$ and $s-$model results are close to each other and they agree with the data within less than 1$\sigma$. 
Results from other calculations, briefly presented in previous sections, show similar trends.
We will come back to those works in the following section.
%
{\squeezetable
\begin{table}[ht]
\caption{\footnotesize The $K^-p$ threshold strong branching ratios, kaonic atom $1s$ level shift and width (in eV).  
See text for experimental data References.
Results for BNW models were obtained using their parameters in our code.}
\begin{center}
\begin{ruledtabular}
\begin{tabular}{lccccc}
Authors [Ref.] & $\gamma$ & $R_c$ & $R_n$ & -$\Delta E_{1s}$ & $\Gamma_{1s}$ \\
\hline
Present work ($c$)& 2.36 & 0.646 & 0.190 & 314& 589  \\
Borasoy {\it et al.}~\cite{Borasoy:2005ie} 
(BNW ($c$))& 2.36 & 0.655 & 0.191 & 316 & 562 \\
Present work ($s$)& 2.40 & 0.645 & 0.189 & 304& 591  \\
Borasoy {\it et al.}~\cite{Borasoy:2005ie} 
(BNW ($s$))& 2.27 & 0.652 & 0.192 & 350 & 535 \\
Mai-Meissner~\cite{Mai:2012dt}  ($c$-type model) & 
2.44$\pm$0.70 & 0.643$\pm$0.017 & 0.268$\pm$0.098 & 296$\pm$52 & 600$\pm$49 \\
Ikeda, Hyodo, Weise~\cite{Ikeda:2012au} ($u$-model)& 
2.37 & 0.66 & 0.19 & 306 & 591 \\
{Shevchenko~\cite{Shevchenko:2012wm} (one-pole)}&  &  &  & 313 & 597 \\
{Shevchenko~\cite{Shevchenko:2012wm} (two-pole)}&  &  &  & 308 & 602 \\
Cieply-Smejkal~\cite{Cieply:2011ig} (NLO)& 2.37 & 0.660 & 0.191 & 310 & 607 \\
Krejcirik~\cite{Krejcirik:2012uw} & 2.36 & 0.637 & 0.178 & 296 & 761 \\
\hline
Experiment& 2.36$\pm$ 0.04 & 0.664$\pm$0.011 & 
0.189$\pm$0.015 & 283$\pm$36 & 541$\pm$92 \\
\end{tabular}
\label{tab:res}
\end{ruledtabular}
\end{center}
\end{table}
}

It is worth recalling that the work by Borasoy {\it et al.}~\cite{Borasoy:2005ie} (BNW) was published before the 
SIDDHARTA data release. 

The only significant deviation from data concerns the width $\Gamma_{1s}$ reported in a very recent
work~\cite{Krejcirik:2012uw}, based on the first order Lagrangian, solving the Lippmann-Schwinger
equation with separable interaction potential.
\subsubsection{Scattering lengths}
\label{sec:sl}
In view of the $K^-d$ scattering length investigations, besides $K^-p$ initiated processes, Eq. (\ref{eq:ch1-8}),
we need to determine amplitudes for the reactions having as initial state $K^-n$, Eq. (\ref{eq:ch1-5}),
for which there are no data. The adopted procedure is then as follows: 
once the parameters for the $T$-matrices for the $K^-p$ channels, Eq. ({\ref{eq:ch1-8}), are determined by fit to 
the data, they are used to calculate the $T$-matrices for the $K^-n$ channels, Eq. (\ref{eq:ch1-5}), by assuming $SU(3)$ 
symmetry in the coupling strengths.   
Those two sets of amplitudes are then used in the three-body calculation of $A_{K^-d}$.

Here along with the corresponding quantity in the $K^-p$ initiated channels, we present the scattering lengths 
as given in Table~\ref{tab:AKN}.
For "Data", kaonic atom measurements are used to extract the scattering lengths.  
"Data" for KEK and DEAR are from Weise~\cite{Weise:2010xn}.
For SIDDHARTA, we used the improved~\cite{Meissner:2004jr} 
Deser-Trueman formula to relate the measured quantities 
to the complex $K^-p \to K^-p$ scattering length
\begin{equation}
\label{eq:D-T}
\Delta E_{1s} + \frac{1}{2} i  \Gamma_{1s} = 2 \alpha^3 \mu^2 a_{K^-p}
\Big [1 - 2 \alpha \mu (ln~\alpha - 1) a_{K^-p} \Big ],
\end{equation}
with $\alpha$ the fine-structure constant and $\mu$ the $K^-$-proton 
reduced mass.
%
%
{\squeezetable
\begin{table}[ht]
\caption{\footnotesize $\Kbar N$ scattering lengths (in fm); $a_p$, $a_n^\circ$ and $a_{ex}$ are calculated 
at $W=M_{K^-}+M_p$ and $a_n$ at $W=M_{K^-}+M_n$.
See text for "Data" explanation.
}
\label{tab:AKN}
\begin{center}
\begin{ruledtabular}
\begin{tabular}{lcccc}
Model &  $a_p(K^-p \to K^-p)$ & $a_n(K^-n \to K^-n)$ & 
$a_n^\circ(\Kbar^\circ n \to \Kbar^\circ n)$  & 
$a_{ex}(K^-p \to \Kbar^\circ n)$  \\  
\hline 
Present work ($c$) & $ -0.72 + i\, 0.90$ & 
$ ~0.86 + i\, 0.71 $  & $ -0.12 + i\, 0.90 $ & $ -1.21 + i\, 0.37 $  \\
Borasoy {\it et al.}~\cite{Borasoy:2005ie}, BNW ($c$)  & 
$ -0.74 + i\, 0.86 $ & $  ~0.61 + i\, 0.71 $  & 
$ -0.24 + i\, 0.96 $ & $ -1.09 + i\, 0.34 $   \\
Present work ($s$)  & $ -0.69 + i\, 0.89 $ & 
$  ~0.90 + i\, 0.66 $  & $ -0.10 + i\, 0.87 $ & $ -1.21 + i\, 0.38 $  \\
Borasoy {\it et al.}~\cite{Borasoy:2005ie}, BNW ($s$) & 
$ -0.85 + i\, 0.86 $ & $ ~0.49 + i\, 0.67 $  & $ -0.38 + i\, 1.01 $ & 
$ -1.11 + i\, 0.36 $ \\
Ikeda {\it et al.}~\cite{Ikeda:2012au} ($u$-model) & 
$ -0.70 + i\, 0.89 $ & $  ~0.57 + i\, 0.73 $  & & \\
Mai-Meissner~\cite{Mai:2012dt}  ($c$-type model) & 
$(-0.68\pm0.15)+ i\, (0.90\pm0.13)$ \\
Shevchenko~\cite{Shevchenko:2012wm} (one-pole)& $-0.76 + i\, 0.89$ \\
Shevchenko~\cite{Shevchenko:2012wm} (two-pole) & $-0.74 + i\, 0.90$ \\
\hline 
"Data": & & & & \\
SIDDHARTA & $( -0.66\pm0.07)+ i\, (0.81\pm0.15)$ & & & \\
KEK & $( -0.78\pm0.18)+ i\, (0.49\pm0.37)$ & & & \\
DEAR & $( -0.47\pm0.10)+ i\, (0.30\pm0.17)$ & & & \\
\end{tabular}
\end{ruledtabular}
\end{center}
\end{table}
}
 
Note that the scattering lengths in Table \ref{tab:AKN} have been obtained at the $K^-p$ threshold (except for 
the elastic $K^-n$ process). 
In fact, these quantities are very sensitive to the value of the threshold at which they are calculated,
which is then reflected in the values obtained for the $A_{K^-d}$ scattering length. 
These aspects have been discussed in a previous paper~\cite{Bahaoui:2003xb}. 

The scattering lengths, Table~\ref{tab:AKN}, show several features as summarized in the following:
{\it (i)} both real and imaginary parts of  $a_p$ agree with data within 1$\sigma$ for all models, with the 
only exception being BNW model-$c$ produced before the release of SIDDHARTA data; 
{\it (ii)} for the three other scattering lengths, models' predictions for the imaginary parts, as well as for 
${\mathcal Re} (a_{ex})$ are compatible with each other;
{\it (iii)} finally, in the case of ${\mathcal Re} (a_n)$ and ${\mathcal Re} (a_n^\circ)$, our predictions turn out 
to be significantly larger than results from other findings. 
The sensitivity of scattering lengths to model ingredients, especially to the NLO contributions, are discussed in 
Ref.~\cite{Ikeda:2012au}.  
%
%
\subsubsection{Scattering amplitudes}
\label{sec:sa}
It is instructive to investigate the real and imaginary parts of the 
$\Kbar N$ scattering amplitudes below 
threshold: $\sqrt{s}\approx$ 1.432 GeV for $K^-p$, 1.433 GeV for $K^-n$ 
and 1.437 GeV for 
$\Kbar ^\circ n$.

Figures~\ref{fig:amp-p} and~\ref{fig:amp-n} show the amplitudes ($f$) for elastic scattering channels,
obtained from our model $c$, as well as those from BNW. An interesting feature is that for $K^-p \to K^-p$ 
both real and imaginary part of the scattering amplitudes are very close, as predicted by the two depicted models.
This however is not the case for the $K^-n \to K^-n$ process, where the real part from our model turns out to be 
significantly larger than the one in BNW, which used DEAR data.

A behavior common to both models is that the maximum of ${\mathcal Im} (f_{K^-p \to K^-p})$ is located at 
$\sqrt{s}\approx$1.416 GeV, so in between the $\Lambda(1405)$ and $\sqrt{s_{K^-p}}\approx$1.432 GeV.

DEAR data were also used in Refs.~\cite{Cieply:2009ea,Hyodo:2007jq,Revai:2008wj}.
A chiral $SU(3)$ coupled-channel dynamics by Hyodo and Weise~\cite{Hyodo:2007jq}, embodying only the 
dominant $\Kbar p -\pi \Sigma$ channels, lead to (much) larger amplitudes, in terms of absolute values.
In a separable meson-baryon approach, Cieply and Smejkal~\cite{Cieply:2009ea}, using a two-pole configuration,
reported larger real part, but comparable imaginary part, though with the maximum around 1.4 GeV.
In a recent work based on a coupled-channels Bethe-Salpeter approach, Mai and Meissner~\cite{Mai:2012dt}
investigated the two-pole structure and found rather narrow spectrum.


\begin{figure}[ht]
\includegraphics[width=0.53\linewidth] {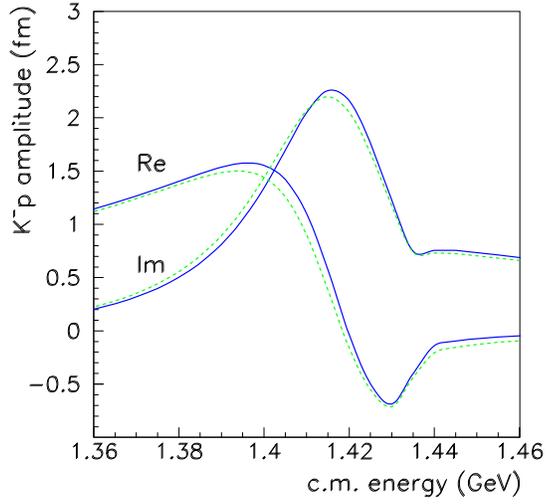} 
\caption{(Color online) \footnotesize 
Scattering amplitudes for $K^-p$ channel within model $c$ with real and imaginary parts 
from the present work (blue solid curves) with $N_{K^-p}=8$, and BNW~\cite{Borasoy:2005ie} 
(green dotted curves) with $N_{K^-p}=10$, obtained using our code.
}
\label{fig:amp-p}
\end{figure}
%
\begin{figure}[hb!]
\includegraphics[width=0.53\linewidth] {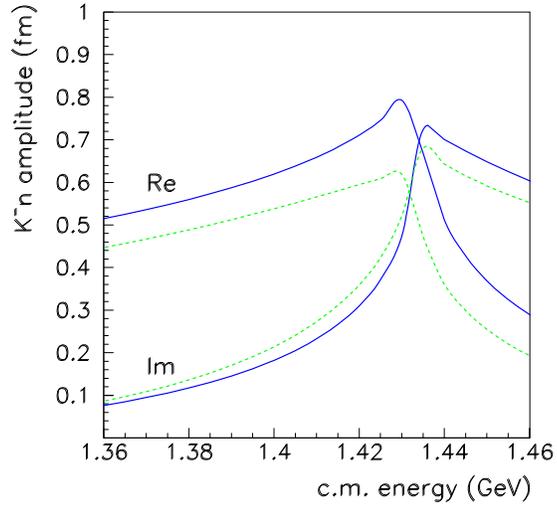}
\caption{(Color online) \footnotesize Same as Fig.~\ref{fig:amp-p}, but for $K^-n$; $N_{K^-n}=5(6)$
for our model (BNW model). 
}
\label{fig:amp-n}
\end{figure}

For the $K^-n \to K^-n$ scattering amplitudes there are fewer predictions available.
In addition to that by BNW, shown in Fig.~\ref{fig:amp-n}, Cieply and Smejkal~\cite{Cieply:2009ea} have
reported amplitudes with ${\mathcal Re} (f_{K^-n \to K^-n})$ roughly 50\% larger than our values and 
${\mathcal Im} (f_{K^-n \to K^-n})$ comparable to our results, but with a narrower structure. 
%
%
\subsubsection{$\pi \Sigma$ invariant-mass observable}
\label{sec:ms}

The $(\pi \Sigma)^\circ$ invariant-mass spectra data were not fitted in the present work. 
So, in Fig.~\ref{fig:spectres} we compare our prediction (full curve) to the $\pi^- \Sigma^+$ 
data~\cite{Hemingway:1984pz} going back to early 80's, measured at CERN in the 
$K^-p \ \to \Sigma^+ \pi^- \pi^+ \pi^-$ reaction at $P_K=4.2$ GeV/c; the agreement is reasonable.
%
\begin{figure}[hb]
\includegraphics[width=0.8\linewidth]{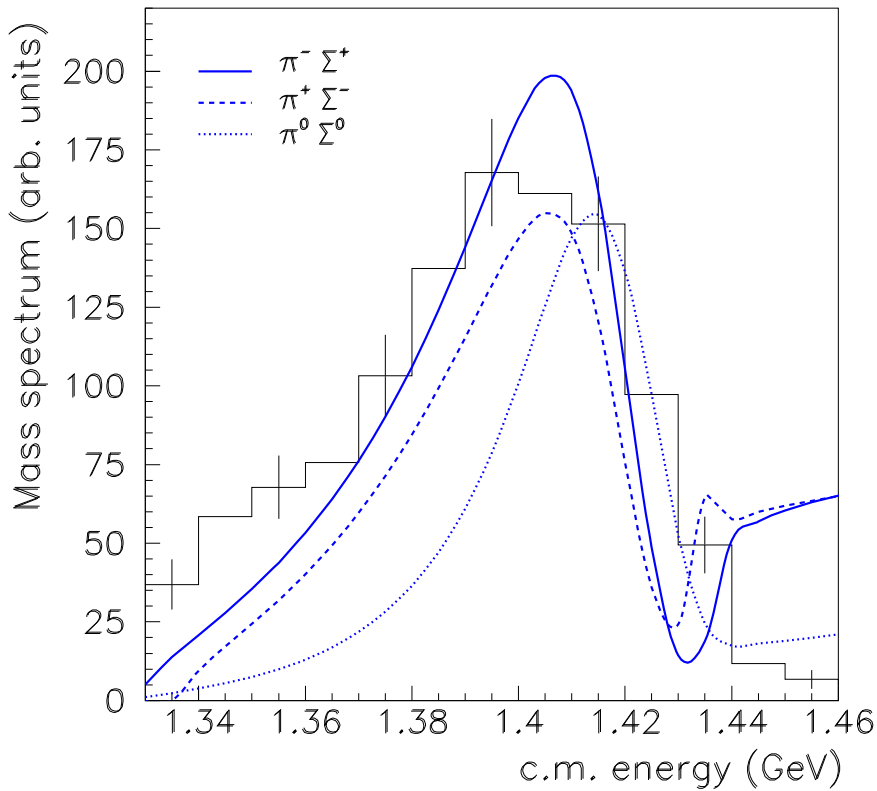}
\caption{\footnotesize 
$(\pi \Sigma)^\circ$ invariant mass spectra for 
$K^- p \to \pi^- \Sigma^+,~ \pi^+ \Sigma^-,~ \pi^\circ \Sigma^\circ$
as a function of the total center-of-mass energy.
}
\label{fig:spectres}
\end{figure}

Given that $\Lambda(1405)$ formation, decaying to $\pi \Sigma$, is about 30 MeV below $K^-p$ threshold, 
a reliable extrapolation of $\Kbar N \to \pi \Sigma$ requires data on the shape and location of the 
invariant mass spectra for $\pi^- \Sigma^+$, $\pi^+ \Sigma^-$ and $\pi^\circ \Sigma^\circ$ channels, for 
which our predictions are depicted in Fig.~\ref{fig:spectres} in full, dashed and dotted curves, respectively.
Predictions for all three channels were also reported within chiral approaches~\cite{Nacher:1998mi,Jido:2010rx}.
It is worth noting that those results endorse our predictions, namely, 
{\it i)} the magnitude of the $\pi^- \Sigma^+$ peak is higher than those of the other channels, which turn 
out to have comparable strengths; 
{\it ii)} the peak of the $\pi^- \Sigma^+$ channel is located at lower energy than those of the two other 
charge states.
 
A recent measurement of the $\pi^\circ \Sigma^\circ$ was performed by the ANKE Collaboration~\cite{Zychor:2007gf}
for $P^{lab}$ = 3.65 GeV/c via the process $pp \to p K^+ \pi^\circ \Sigma^\circ$ with rather large 
uncertainties, as discussed in~\cite{Esmaili:2009rf}, and successfully reproduced within a dynamical chiral 
$SU(3)$ calculation~\cite{Geng:2007vm}. 
Very recently, HADES collaboration released data~\cite{Agakishiev:2012qx,Agakishiev:2012xk} for 
$p p \to \pi^\pm \Sigma^\mp K^+ p$ at $E^{lab}_{kin}$ = 3.5 GeV, 
and $\Lambda(1405) \to \pi^\pm \Sigma^\mp $ channels were extracted. 
To that purpose, and in order to subtract contributions from 
$p p \to  \Sigma^\circ(1385) K^+ p \to \pi^\pm \Sigma^\mp K^+ p$, the experimentally determined~\cite{Epp12} 
cross sections ratio $\sigma_{p p \to  \Sigma^\circ(1385) K^+ p}/\sigma_{p p \to  \Lambda(1405) K^+ p}$
was used.

HADES collaboration results~\cite{Agakishiev:2012xk} lead to mass spectra different from the rather old kaon 
beam data~\cite{Hemingway:1984pz} depicted in Fig.~\ref{fig:spectres}, but also from preliminary measurements with 
photon beams at JLab~\cite{Moriya:2011af} and SPring-8/LEPS~\cite{Ahn:2010zzb}. 
Final results using electromagnetic probes will hopefully allow checking if the mass spectra depend on the 
entrance channel, as claimed by the HADES collaboration~\cite{Agakishiev:2012xk}.

Moreover, experimental project at J-PARC~\cite{Enomoto:2011zz} on in-flight ($K^-,n$) reaction on deuteron is 
expected to deepen our understanding of the $(\pi \Sigma)^\circ$ invariant-mass spectra, via the
$K^-d \to (\pi \Sigma)^\circ n$ process~\cite{Miyagawa:2012xz,Jido:2012cy}.    

Finally, in order to get insights into the quasibound $K^-p$ states, {\it via} $\pi \Sigma$ mass spectra, 
$K^-$ absorption in deuterium~\cite{Esmaili:2009rf,Jido:2010rx,Kopeliovich:2011zzc}, as well as in 
$^3He$ and $^4He$~\cite{Yamazaki:2002uh,Akaishi:2002bg,Esmaili:2009iq} have been investigated theoretically.
For a recent review on the nature of $\Lambda(1405)$ see Ref.~\cite{Hyodo:2011ur}. 
\subsubsection{Summary on two-body interactions}
\label{sec:sumKN} 
The main novelty of the present work with respect to our previous investigations~\cite{Bahaoui:2002yc,Bahaoui:2003xb}
is the chiral unitary approach to describe $\Kbar N$ interactions, as presented above.

In order to facilitate the reading of the paper, we summarize the ingredients for other two-body channels needed 
to move to the $K^-d$ studies, as reported in detail in Ref.~\cite{Bahaoui:2003xb}.

{\bf a) $NN$ interactions:}

For the deuteron ($d$) channel, we used a relativistic separable potential. 
All the interactions considered are of rank-1, which allows correctly reproducing the static parameters; namely, 
the triplet effective range parameters $a_t$ and $r_t$, the $D$-state percentage value $P_D$, the quadrupole moment $Q$, 
and the asymptotic ratio $\eta=A_D/A_S$, the $^3S_1$ phase shift, and also the deuteron monopole charge form factor 
up to about 6 {\it fm}$^{-1}$.  
In this model the $D$-state probability is $P_D=6.7 \%$. 
All details can be found in Ref.~\cite{Giraud:1979tj}.

{\bf b) $\pi N-P_{33}$ and hyperon-proton interactions:}  

Using the same models as in Ref.~\cite{Bahaoui:2003xb}, we have found that these interactions do not have a 
significant effect on the $K^- d$ scattering length. 
We therefore will not bring these partial waves in our discussion.

In conclusion, the obtained $\Kbar N$ model and handling of the $NN$ two-body channels lead to a reliable enough 
elementary interactions description to be used in the three-body calculation. 
  
%
%
\section{$K^- d$ scattering length}
\label{sec:Kd}
Pioneer work by Toker, Gal and Eisenberg~\cite{Toker:1981zh}, some three decades ago, initiated theoretical
investigations~\cite{Bahaoui:2002yc,Bahaoui:2003xb,Meissner:2006gx,Faber:2010iw,Barrett:1999cw,Deloff:1999gc,
Grishina:2004rd,Ivanov:2004bv} 
on the  $K^-d$ scattering length, in spite of lack of data, though foreseen in a near future.
For recent reviews see Refs.~\cite{Gal:2006cw,Gasser:2007zt}.
\subsection{Three-body equations for the $K^- d$ system}
\label{sec:theory}
Here, we summarize the three-body equations formalism for the $K^- d$ system, in which the two-body 
input described in the previous Section will enter. 
Then we give the expression for the  $K^- d$ scattering length.

As the calculations are performed in particle basis, the equations below are given in this basis. 

The three-body formalism, where the two-body operators connect two states embodying particles which 
are different, lead to the following system of coupled equations, written in operator form:

\begin{equation}
\label{eq:3body}
X_{a\,  b}(s) = Z_{a\, b}(s) + \sum_{c,\,c'} Z_{a\, c}(s) 
                                  R_{c\, c'}(s)\; X_{c'\, b}(s) ,
\end{equation}  
with $s$ the three-body total energy.
Here, $a,b,c$ and $c'$ are the indices which specify the particles involved in the spectator and the interacting pair 
three-body channels. 
$X_{ab}$ is the transition amplitude between channels $a$ and $b$, and  $Z_{ab}$ is the corresponding Born term.
The two-body operator $R_{c\, c'}$ connects two different two-body states labeled as $c$ and $c'$. 

In the particle basis the particles belonging to the various isospin multiplets do have their physical masses.
The number of 3-body channels is thus considerably enlarged compared to the case of isospin basis. So, when 
using our $c$- or $s$- model, the $\Kbar N$ interactions must include the 8 channels coupled to $K^-p$, namely: 
$K^-p$, $\Kbar^\circ n$, $\pi^-\Sigma^+$, $\pi^+\Sigma^-$, $\pi^\circ\Sigma^\circ$,
$\pi^\circ\Lambda$, $\eta\Sigma^\circ$, $\eta\Lambda$,
as well as the 5 channels coupled to $K^-n$:
$K^-n$, $\pi^-\Sigma^\circ$, $\pi^\circ\Sigma^-$, $\pi^-\Lambda$, $\eta\Sigma^-$.

Now, we specify the values taken by the channel indices in Eq.~(\ref{eq:3body}). 
Taking into account the deuteron and the 13 ($N_{K^-p}+N_{K^-n}=8+5$) two-body inputs in our approach, one must 
consider the three-body channels in the particle basis as given in Table~\ref{tab:labels}, where in the first 
and fourth lines the {\it spectator} particles followed (in parenthesis) by the associated {\it pair} are specified.

%
ù
\begin{table}[h]
\caption{\footnotesize The three-body channels in the particle basis. 
The second line specifies the isospin of the two-body sub-system and the third one the label.}
\label{tab:labels}
\begin{center}
\begin{tabular}{lccccccc}
\hline
\hline
Channel & $K^-(pn)$ & $n(K^-p)$ & $n(\Kbar^\circ n)$ & 
$n(\pi^-\Sigma^+)$ & $n(\pi^+\Sigma^-)$ & $n(\pi^\circ \Sigma^\circ)$ & 
$n(\pi^\circ \Lambda)$ \\
\hline
Isospin & 0 & 0 & 0 & 0 & 0 & 0 & 0  \\
Label & $d$ & $y_1$ & $y_2$ & $\alpha_1$ & $\alpha_2$ & $\alpha_3$ & 
$\alpha_4$  \\	
\hline
\hline
Channel  & $n(\eta\Sigma^\circ)$ & $n(\eta\Lambda)$ &	$p(K^-n)$ &
	        $p(\pi^-\Sigma^\circ)$ & $p(\pi^\circ\Sigma^-)$ &	
		$p(\pi^-\Lambda) $ & $p(\eta\Sigma^-) $  \\
\hline
Isospin & 0 & 0 & 1 & 1 & 1 & 1 & 1  \\
Label & $\mu_1$ & $\mu_2$ & $y_3$ & $\alpha_5$ & $\alpha_6$ & $\alpha_7$ & 
$\mu_3$ \\
\hline
\hline
\end{tabular}
\end{center}
\end{table}
%

Starting from the formal equations (\ref{eq:3body}), the final relativistic equations for the rotationally 
invariant amplitudes are derived~\cite{Bahaoui:2003xb,Aaron:1969my,Rinat:1976qg,Bah_thesis}:
\begin{eqnarray}
\label{eq:rotinv}
X_{\tau_a \tau_c}^{\cal J}(p_a , p_c ; s) &=
& Z_{\tau_a \tau_c}^{\cal J}(p_a , p_c ; s) \nonumber\\
&+& \sum_{b,\tau_b ; b',\tau_{b'}} \int\frac{dp_b \, p_b^2}{2\epsilon_{b}}
   Z_{\tau_a \tau_b}^{\cal J}(p_a , p_b ; s)R_{bb'}^{c_b=c_{b'}}(\sigma_{b})
   X_{\tau_{b'} \tau_c}^{\cal J}(p_b , p_c ; s)\,,
\end{eqnarray} 
where $\sigma_{b}$ is the invariant energy of the pair in channel $b$ expressed in the three-body 
center of mass system, $c_a = (J_a, S_a, I_a)$ specifies the conserved quantum numbers of the
pair in channel $a$, and $\tau_a=(c_a, l_a,\Sigma_a)$ specifies the three-body quantum numbers in 
this channel. Labels $c$ and $\tau$ refer to the spin-isospin variables in a given channel.
For example, assuming that channel $a$ is composed with particle $i$ as spectator and the pair $(jk)$, 
we define the following quantities:
\begin{itemize}
\item $\bm s_i$: spin of particle $i$,
\item $\bm S_i$ $(= \bm s_j + \bm s_k)$, $\bm L_i$, and 
$\bm J_i$ $(= \bm L_i + \bm S_i)$: spin, 
      orbital angular momentum, and total angular momentum, respectively, 
      of pair $(jk)$,
\item $\bm \Sigma_i$ $( = \bm s_i + \bm J_i)$, $\bm l_i$, and 
$ \cal J$ $(= \bm l_i + \bm \Sigma_i)$:  
      channel spin, orbital angular momentum of $i$ and $(jk)$, 
      and  three-body total angular momentum, 
      respectively.
\end{itemize}
The Born terms matrix in the considered case is a $14 \times 14$ matrix. 
However, only a few terms are non-zero, as explained in what follows:

{\it (i)} $Z_p = \bra{K^-(pn)} G_0 \ket{n(K^-p)}$: exchange of the $p$ between the deuteron and the 
$(K^-p)$ pair,

{\it (ii)} $Z_n = \bra{K^-(pn)} G_0 \ket{p(K^-n)}$: exchange of the $n$ between the deuteron and the 
$(K^-n)$ pair, 

{\it (iii)} $Z_{K^-} = \bra{n(K^-p)} G_0 \ket{p(K^-n)}$: exchange of the $K^-$ between the $(K^-p)$ 
and $(K^-n)$ pairs,

{\it (iv)} $Z_{\Kbar^\circ} = 
\bra{n(\Kbar^\circ n)} G_0 \ket{n(\Kbar^\circ n)}$: exchange of the $\Kbar^\circ$ between the 
$(\Kbar^\circ n)$ pairs

Of course, we must add the corresponding symmetric Born terms, for which the explicit expressions can be 
found in Ref.~\cite{Bah_thesis}. 

In order to obtain the rotationally invariant equations, we only have to antisymmetrize the 
$Z_{\Kbar^\circ} = \bra{n(\Kbar^\circ n)} G_0 \ket{n(\Kbar^\circ n)}$ Born term. 
This is done as described in Appendix C of Ref.~\cite{Bahaoui:2003xb}. 

Concerning the $\Kbar N$ two-body propagators, they are evaluated as described in Sec. II and Appendix A 
of the present paper, and the $d$ propagator as explained in Ref.~\cite{Bahaoui:2003xb}. 
%
%
\subsection{\bf Practical calculation}
\label{sec:practical-cal} 
To end this Section, we consider the $K^-d$ scattering length, defined as:
\begin{equation}
A_{K^-d}= - \displaystyle\lim_{P_{K} \to 0} \frac{1}{32\pi^2\sqrt s} X_{dd} ,
\end{equation}
where $X_{dd}$ is the $({\cal J}=1^-, l=l'=0)$ partial amplitude for $K^-d$ elastic scattering, 
evaluated at the zero limit for the kaon momentum.

If we retain the contributions of the $d$+$\Kbar N$ two-body channels, we have a system of 14 coupled 
three-body channels (see Table \ref{tab:labels}). 
After angular momentum reduction, we obtain (in the particle basis) a system of 28 coupled equations 
for ${\cal J}=1^-$, when including the deuteron $d$, the 8 channels coupled to $K^-p$, and the 5 channels 
coupled to $K^-n$; see Table \ref{tab:qn23c}. 

The singularities of the kernel are avoided by using the rotated contour method~\cite{Hetherington:1967zza}, 
and, after discretization of the integrals, this system is transformed into a system of linear equations.
In order to solve this system, we use the Pad\'e approximants technique which leads to a convergent solution 
from the successive iterated terms. 
In practice, we have used a diagonal $[5/5]$ Pad\'e (constructed with the 11 first iterates), which was found to 
be sufficient to achieve convergence. 
Even if the dimension of the matrix to be inverted is rather large, this is a sparse matrix  because of the 
limited number of non-zero Born terms, and the Pad\'e approximants method is much less time consuming to solve 
the linear system than the usual matrix inversion method.
%
\squeezetable
\begin{table}[h]
\caption{\footnotesize Two-body ($L,S,J,T$) and three-body ($l,\Sigma,\cal J$)  
quantum numbers in the particle basis, in the case ${\cal J}=1^-$, for
$N_{K^-p}=8$ and $N_{K^-n}=5$.} 
\label{tab:qn23c}
\begin{center}
\begin{ruledtabular}
\begin{tabular}{cccccccccc}
 channel   & & name   & $L$ & $S$ & $J$ & $T$ & & $\Sigma$ & $l$   \\
\hline
$K^-(np)$ & & $d$  & 0,2 & 1 & 1 &  0 &  & 1 & ${\cal J}+1$       \\
         &    &     &     &   &   &    &  &   & ${\cal J}-1$       \\
\hline
$n(K^- p,\ \Kbar^\circ n,\ \pi^-\Sigma^+,\ \pi^+\Sigma^-,
\ \pi^\circ\Sigma^\circ,
\ \pi^\circ\Lambda,\ \eta\Sigma^\circ,\ \eta\Lambda)$  
   & & $y_1,y_2,\alpha_1,\alpha_2,\alpha_3,\alpha_4,\mu_1,\mu_2$ 
                      & 0 & 1/2 & 1/2 & 0 &  & 1 &  ${\cal J}+1$    \\
              &       &   &     &     &   &  &  & & ${\cal J}-1$    \\
\hline
$p(K^-n,\ \pi^-\Sigma^\circ,\ \pi^\circ\Sigma^-\ ,
\pi^-\Lambda,\ \eta \Sigma^-)$  
   &  & $y_3,\alpha_5,\alpha_6,\alpha_7,\mu_3$
                      & 0 & 1/2 & 1/2 & 1 &  & 1 &  ${\cal J}+1$     \\
              &       &   &     &     &   &  &  & &  ${\cal J}-1$    
\end{tabular}
\end{ruledtabular}
\end{center}
\end{table}
%
%
\subsection{Results and Discussion}     
\label{sec:results}
The zero limit for the kaon momentum corresponds to $W \equiv \sqrt s$=1431.95 MeV.
As previously emphasized~\cite{Bahaoui:2003xb}, 
the ${\Kbar N}$ scattering length, and in consequence $A_{K^-d}$, show strong dependence on W, due to 
proximity of the dominant $\Lambda$(1405).
It is worth to note that the full three-body results are not subject to excessive sensitivities. 
Actually, the energy dependence in the vicinity of zero limit for the kaon momentum is smeared out due 
to a loop momentum integral.

Our results for $A_{K^-d}$, and those from other recent works, are reported in Table~\ref{tab:AKd}.
We observe that the $c$- and $s$-models for ${\Kbar N}$ lead to almost identical values for $A_{K^-d}$.
%
\begin{table}[htb]
\caption{\footnotesize $K^-d$ scattering length (in fm).}
\begin{center}
\begin{tabular}{lccc}
\hline
\hline
Authors [Ref.] &&& $A_{K^-d}$  \\
\hline
Present work ($c$)&&& -1.58 + i\, 1.37   \\
Borasoy {\it et al.}~\cite{Borasoy:2005ie} (BNW ($c$))&&& -1.59 + i\, 1.59  \\
Present work ($s$)&&& -1.57 + i\, 1.37   \\
Borasoy {\it et al.}~\cite{Borasoy:2005ie} (BNW ($s$))&&& -1.67 + i\, 1.52  \\
Doring-Meissner~\cite{Doring:2011xc} &&& -1.46 + i\, 1.08 \\
Shevchenko~\cite{Shevchenko:2012wm} (one-pole) &&& -1.48 + i\, 1.22 \\
Shevchenko~\cite{Shevchenko:2012wm} (two-pole) &&& -1.51 + i\, 1.23 \\
Revai~\cite{Revai:2012fx}  (one-pole) &&& -1.52 + i\, 0.98 \\
Revai~\cite{Revai:2012fx}  (two-pole) &&& -1.60 + i\, 1.12 \\
Oset {\it et al.}~\cite{Oset:2012gi}   &&& -1.54 + i\, 1.82 \\
Bahaoui {\it et al.}~\cite{Bahaoui:2003xb} &&& -1.80 + i\, 1.55 \\
\hline
\hline
\end{tabular}
\label{tab:AKd}
\end{center}
\end{table}

A first comment concerns the fact that our present results, both for real and imaginary parts 
of $A_{K^-d}$, come out about 12\% smaller in magnitude with respect to our 2003 values~\cite{Bahaoui:2003xb}.
These changes are due on the one hand to the improved $\Kbar N$ model presented in Sec.~\ref{sec:two-body},
and on the other hand to the since then published SIDDHARTA data~\cite{Bazzi:2011zj}. 
With respect to numerical procedure in $\Kbar N$ T-matrices calculations, we have carefully checked the 
convergence of the integral. 
In our previous model it was ensured by a form factor in the separable form of the interaction, while in 
the present approach, with no explicit form factor, the amplitudes are tempered by dimensional regularization and 
converge, though more slowly. 

Works by other authors (Table~\ref{tab:AKd}) show also the impact of the model used for $\Kbar N$ interactions 
and the inclusion or not of the SIDDHARTA data, as discussed below.

Comparing our results with those of BNW, performed before SIDDHARTA data became available, we observe that: 
{\it (i)} Real parts for model $c$ are identical, while for the $s$-model BNW values are in between our present 
and past~\cite{Bahaoui:2003xb} results.
So, the $c$-model is less affected than the $s$-model by the SIDDHARTA data.
{\it (ii)} The imaginary parts, for both $c$- and $s$-models in BNW are closer to our previous results rather than 
to those reported in this paper, underlining the significant constraints brought in by the SIDDHARTA data.

All recent results~\cite{Doring:2011xc,Shevchenko:2012wm,Revai:2012fx,Oset:2012gi} reported in
Table~\ref{tab:AKd} were obtained considering the SIDDHARTA data, except in Ref.~\cite{Revai:2012fx}. 

A new parameterization of the $\Kbar N \to \pi \Sigma$ potentials was successfully performed by 
Shevchenko~\cite{Shevchenko:2012wm} and embodied in the author's~\cite{Shevchenko:2011ce} coupled-channels 
Faddeev equations in Alt-Grassberger-Sandhas form. 
Results for one- and two-pole structure of $\Lambda$(1405) lead to almost identical values (Table~\ref{tab:AKd})
for $A_{K^-d}$, showing that the scattering length is not sensitive enough to the poles structure of the 
$\Lambda$(1405) resonance. 
Both real and imaginary parts are comparable to our values within less than 10\%.
Note that in Ref.~\cite{Shevchenko:2012wm} the $NN$ interactions contain no $D$-state. 

Revai~\cite{Revai:2012fx} using a similar three-body formalism, finds slightly larger differences between one- 
and two-pole schemes, especially for the imaginary part of the scattering length, which turns out be lower 
than predictions from all other works quoted in Table~\ref{tab:AKd}. 
These features might be due to the fact that the two-body channels have been investigated using the KEK data 
rather than the SIDDHARTA results. Actually, Shevchenko has reported such a sensitivity considering 
SIDDHARTA~\cite{Shevchenko:2012wm} versus KEK~\cite{Shevchenko:2011ce} measurements. 

Finally, two recent works based on the fixed center approximation (FCA), within a non-relativistic 
effective field theory~\cite{Doring:2011xc} and Faddeev equations~\cite{Oset:2012gi}, lead to discrepancies 
in real and imaginary parts of about 24\% and 44\%, respectively. 
However, a careful study~\cite{Doring:2011xc} on the accuracy of predicted $A_{K^-d}$ puts forward an "allowed" 
surface in the [${\mathcal Re} (A_{K^-d})$, ${\mathcal Im} (A_{K^-d})$]-plane. 
Interestingly, all the results shown in Table~\ref{tab:AKd} fall within a rather compact sub-space of that plane.
The forthcoming data from the SIDDHARTA Collaboration~\cite{Okada:2010zz} will, hopefully, provide sufficient 
constraints on that physical sub-space.   
%
%
\section{Summary and Conclusions} 
\label{sec:conclu}
In section~\ref{sec:two-body} we performed a comprehensive study on the two-body $\Kbar N$ interactions at low 
energies via a Chiral Unitary Model including the next to the leading order terms and the coupled Bethe-Salpeter
equations. 
The emphasize was put on the recent kaonic hydrogen data released bythe SIDDHARTA Collaboration~\cite{Bazzi:2011zj}, 
which allowed a consistent use of low energy data to extract the 13 adjustable parameters in our approach and 
reproduce satisfactorily the fitted data.

Using SU(3) symmetry and the scattering amplitudes for the processes
$K^-~p~\to~K^-p,~ \Kbar^\circ n,~\Lambda \pi^\circ,~\Sigma^+ \pi^-,
~\Sigma^\circ \pi^\circ,~\Sigma^-\pi^+,~\Lambda\eta,~\Sigma^\circ\eta$,
those for the reactions $K^- n~\to~K^- n,~\Lambda\pi^-,~\Sigma^\circ\pi^-,
~\Sigma^-\pi^\circ,~\Sigma^-\eta$ were also determined, providing the needed complete $\Kbar N$ inputs 
to the $K^-d$ system investigations.

Within the two-body studies, we put forward predictions for entities of interest, namely, scattering 
amplitudes and lengths, as well as the $(\pi \Sigma)^\circ$ invariant-mass spectra. 
Results from other authors having used the SIDDHARTA data give comparable results e.g. for the scattering length 
$a_{K^-p \to K^-p}$.
Data for $(\pi \Sigma)^\circ$ invariant-mass spectra has never been used to constraint the models, because of 
poor statistics. 
Forthcoming data, hopefully accurate enough, using hadronic~\cite{Zychor:2007gf,Agakishiev:2012qx,Agakishiev:2012xk} 
or electromagnetic probes~\cite{Moriya:2011af,Ahn:2010zzb} are expected to significantly improve our knowledge on the 
$\Lambda(1405) \to \ (\pi \Sigma)^\circ$ transitions and hence on the $\Kbar N$ low energy interactions. 
This latter is also of paramount importance in pinning down whether the antikaon is or not bound in few-body nuclear 
systems, for which advanced formalisms have been developed~\cite{Hyodo:2012pn}.

The elementary operators obtained in section~\ref{sec:two-body} were then used in our three-body equations approach,
embodying the relevant $NN$ two-body channel, namely the deuteron, which was described using a relativistic separable 
potential, with the $D$-state probability $P_D=6.7 \%$. 

Our prediction for the $K^- d$ scattering length is $A_{K^- d}$ = -1.58 + i 1.37. 
Results from various calculations need to be compared to the forthcoming data from the SIDDHARTA 
Collaboration~\cite{Okada:2010zz}. 
%
%
\begin{acknowledgments}
T. M. would like to acknowledge  several pleasant visits extended to him 
at the Thomas Jefferson National 
Accelerator Facility.  He also would like to thank Bugra Borasoy, 
Wolfram Weise, and  Antonio Oller for 
clarifications regarding their works. 
Thanks are also to Tony Thomas for reminding him of the earlier $SU(3)$ 
{\it cloudy bag} approach to the
$\Kbar N - \pi \Sigma$ channels. 
The work of K.T. was supported by the University of Adelaide 
and the Australian Research Council through 
grant No. FL0992247 (AWT). He also would like to acknowledge 
the International Institute of Physics,
Federal University of Rio Grande do Norte, Natal, Brazil,
for warm hospitality during which part of this work
was carried out.
\end{acknowledgments} 
%

\newpage

\begin{appendix}
%
\section{Two-body $\Kbar N$  amplitudes}
\label{apdx:A}
Since we have closely followed the definition and convention used in BNW~\cite{Borasoy:2005ie},  
not much may be needed to repeat the description of the quantities in the $\Kbar N$ equations.   
So, we will simply give some supplementary information.

{\bf (i) Relation to our previous notations}

In case the reader may find it useful, the quantities we adopted in our 2003 work~\cite{Bahaoui:2003xb} 
denoted as {\it LYS $\equiv$ Lyon-Saclay} are related to those in BNW as
\begin{eqnarray}
V_{LYS} &=& - 4\pi V_{BNW},\\
T_{LYS} &=& - 4\pi T_{BNW},\\
G_{LYS} &=& \frac{1}{4\pi} G_{BNW},
\end{eqnarray}
and the corresponding scattering amplitude reads
\begin{equation}
f_{LYS}=-\frac{1}{32\pi^2\sqrt{s}}T_{LYS},
\end{equation}
where $s$ is the Mandelstam variable of the channel.

{\bf (ii) Two-body meson-baryon scalar loop}

In BNW this function is explicitly given in their Eq.~(11).  
We also used the same subtraction scale value, $\mu=1.0$ GeV, and $m$ and $M$  are the generic meson 
and baryon masses, respectively.
For our practical implementation, we have used a somewhat different form of the propagator which 
is completely equivalent to the form in BNW.  
We also checked the equivalence of our form with the one used in Ref.~\cite{Oller:2006jw}.  
For convenience, we split $G$ into three parts:
\begin{equation}
G_{BNW}(s)=G_1+G_2+G_3,
\end{equation}
where the constant term is
\begin{equation}
G_1 =a(\mu)+\frac{1}{32\pi^2}\left [\ln (\frac{m^2}{\mu^2}) 
+ \ln (\frac{M^2}{\mu^2}) -2 \right ],
\end{equation}
with $a(\mu)$ the subtraction constant. 
Then
\begin{equation}
G_2=\frac{1}{32\pi^2 s}(m^2-M^2) \ln (\frac{m^2}{M^2}),
\end{equation}
and
\begin{equation}
G_3=\frac{1}{32\pi^2 s}4mM\sqrt{z^2-1}\ln{(-z-\sqrt{z^2-1})},
\end{equation}
where 
\begin{equation}
z=\frac{s-M^2-m^2}{2mM}.
\end{equation}
  
It might be useful to remind that at the threshold: $s =(M+m)^2$, $z=1$.  
Also $z=-1$ at $s =(M-m)^2$.
Then it can be shown that $G$ is analytic in the whole complex $s$ plane with an unitarity 
cut on the positive real axis starting from the threshold: $s=(M+m)^2$. 
An apparent singularity at $s=0$ in the above expression, which might derive from the 
kinematics at origin, must be absent.
Explicitly we looked at the small $|s|$ behavior and found that
\begin{eqnarray}
G_2 + G_3&\approx&\frac{1}{16 \pi^2} 
         \left[1 - \frac{M^2+m^2}{2 (M^2-m^2)} \ln (\frac{m^2}{M^2}) \right]
\\
&&- \frac{1}{32 \pi^2 (M^2-m^2)^2} 
 \left[ (M^2+m^2) + \frac{2 (Mm)^2}{(M^2-m^2)} \ln (\frac{m^2}{M^2}) \right] s.
\end{eqnarray}

So, in numerical calculation in the three-body code where the vanishing value of $s$ may be encountered, 
those two contributions need to be calculated together, in order to avoid possible spurious divergences.  

\end{appendix}

\newpage


%


\end{document}